\documentclass[fleqn,usenatbib]{mnras}

\usepackage{newtxtext,newtxmath}

\usepackage[T1]{fontenc}
\usepackage{ae,aecompl}

\usepackage{graphicx}	
\usepackage{amsmath}	
\usepackage{amssymb}	

\title[Evidence of the HCB Great Wall]{Re-Examining the Evidence of the Hercules-Corona-Borealis Great Wall}

\author[S. Christian]{
Sam Christian$^{1}$\thanks{E-mail: Samchristianemail@gmail.com}
\\
$^{1}$The Liberal Arts and Science Academy, Austin, Texas\\
}

\date{Accepted XXX. Received YYY; in original form ZZZ}

\pubyear{2020}

\begin{document}
\label{firstpage}
\pagerange{\pageref{firstpage}--\pageref{lastpage}}
\maketitle

\begin{abstract}
In the $\Lambda$-CDM paradigm of cosmology, anisotropies larger than 260 Mpc shouldn't exist. However, the existence of the Hercules-Corona Borealis Great Wall (HCB) is purported to challenge this principle by some with an estimated size exceeding 2000 Mpc. Recently, some have challenged the assertion of the existence of the HCB, attributing the anisotropy to sky exposure effects. It has never been explained why the original methods purporting the existence of the HCB produce anisotropies, even if sky-exposure effects are taken into account. In this paper, I apply the methods of the original papers purporting the existence of the HCB in various Monte-Carlo simulations that assume isotropy to analyze the empirical meaning of the significance levels of the original tests used. I find that, although the statistical tests at first glance show significant anisotropies present in the suspect sample, Monte-Carlo simulations can easily reproduce the sample in most cases, and if not, the differences can be accounted for by other statistical considerations. An updated sample raises the probability of drawing the observed clustering from an isotropic sample ten-fold in some cases. Thus the statistical tests used in prior studies overestimate the significance of the observed anisotropy, and an updated sample returns even less significant probabilities. Given the ability to reproduce the observed anisotropy in Monte-Carlo simulations, the new, higher probabilities of being drawn from isotropy for an updated sample, and the work of previous papers attributing anisotropies to sky-selection effects, the existence of the HCB must be treated as doubtful at best.
\end{abstract}




\section{Introduction}

The large-scale isotropy of the universe is a critical component of the $\Lambda$CDM cosmological model. In the $\Lambda$CDM paradigm, small-scale anisotropies will exist due to quantum fluctuations during inflation, but large-scale isotropy is preserved. Analysis of the cosmic background radiation has backed up this result, with no large-scale anisotropies being found \citep{planck2015, liddle2000}

However, some structures have been found that violate the principle of large-scale isotropy, perhaps the most famous being the Sloan Great Wall. Using the \citet{yadav2010} value of large-scale isotropy being preserved starting at 260 Mpc $h^{-1}$, at least three large quasar groups to date violate large-scale anisotropy as well. The largest of these is the Huge-LQC of \citet{Clowes2013}, although the violation of homogeneity due to the latter has been debated \citep{Nadathur2013}.

Gamma ray bursts (GRBs) have long been found to be isotropic. Indeed, the isotropic distribution of GRBs was one factor that led to the proposal of GRBs being of extragalactic origin  \citep{briggs1996}. With the advent of redshfit measurements of GRBs, this extragalactic origin has been confirmed. Although the isotropy of long-duration GRBs ($T_{90} \geq 2$) has typically been established (see the introduction of \citet{Ripa2019} for a discussion of the history of this), short-duration GRBs have been found to follow an anisotropic distribution \citep{tarnopolski2017, vavrek2008}. As short-duration GRBs have low redshift, their distribution is generally agreed not to violate the cosmological principle, although some have said otherwise \citep{balazs2009}. 

There is a general consensus that the origin of long GRBs is type Ic supernovae \citep{sobacchi2017}. Thus there is a linkage between long GRBs and a young, massive stellar population. If the number density of young, massive stars is assumed to be representative of the overall number density of stars and, more broadly, the density of matter, then long GRBs can be used to trace large-scale structures. Such structures could alternatively represent an area of increased star formation \citep{Balazas2015}.

Recently, anisotropies in long GRBs have been found. The most notable of these is the Hercules-Corona Borealis Great Wall (HCB) between redshifts 1.6 and 2.1 \citep{horvath2014}, which is also the largest identified structure in the universe. The HCB is purported to be larger than allowed by the $\Lambda$-CDM model, although various arguments have been made to explain its existence, particularly a changing scale factor of homogeneity \citep{li2015} and alternatively, the HCB being an area of increased star formation \citep{Balazas2015}. The existence of the HCB is criticized in \citet{ukwatta2016}, which uses kernel-density methods to argue that the observed anisotropies in the redshift dependent clustering can be explained solely by extinction/exposure biases in the data. But \citet{horvath2014} includes a rudimentary bias correction chi-squared test, finding that selection bias can't explain the observed clustering.

Thus, there seems to be no consensus on whether the HCB exists, although in many papers it is still cited as a violation of homogeneity \citep{racz2017, eingorn2017}.

In this paper, I aim to settle the controversy over the existence of the HCB by following the methods of \citet{horvath2014} as closely as possible, only deviating when more precise estimates can be generated. I employ the two-dimensional Kolmogorov-Smirnov test, the nearest-neighbor test, and the bootstrap point radius test as done in \citet{horvath2014}. I then use Monte-Carlo simulations, combined with a consideration of extinction effects, to explain the suspect values generated in \citet{horvath2014}. I also apply these tests to an updated sample of GRBs, as a long time span has passed since the publication of \citet{horvath2015}.

\section{Data Set}

Following the attempt to mirror the methods of \citet{horvath2014}, I use the data set provided in \citet{horvath2015}, an updated study of the HCB using new data since the original study was published. \citet{horvath2015} uses a composite set of GRBs from multiple instruments, although most of the GRBs are from the Swift Gamma Ray Telescope \citep{Sakamoto2008}. The dataset consists of 361 GRBs, with redshifts ranging from 0.0065 to 9.4.

Since it has been four years since \citet{horvath2015} was published, many new GRBs with redshift measurements have occured, so a more complete sample can be tested. For this updated data set, I use the database located at \url{http://www.mpe.mpg.de/~jcg/grbgen.html}.

All code used to generate figures and statistics in this paper is available at \hyperlink{https://github.com/Sam-2727/Gamma-ray-burst-isotropy}{https://github.com/Sam-2727/Gamma-ray-burst-isotropy}. The Github repository also contains a database of the GRBs used in the statistical tests.

In \citet{horvath2015}, the data set of 360 GRBs was divided into four to nine zones of equal numbers of GRBs, and the anisotropy was found to be approximately between redshifts 1.6 and 2.1, equivalent to group two in the four group case, where groups are in increasing order of redshift. This is the data set I discuss in Sec. \ref{sec:significance}.

With my updated data set (Sec. \ref{sec:updated}), I use the original suspect range of \citet{horvath2014} and \citet{horvath2015}.
\section{Tests of Significance}
\label{sec:significance}
\subsection{Statistical Considerations}
\label{sec:statisticalconsiderations}
\begin{figure}
    \centering
    \includegraphics[width=\columnwidth]{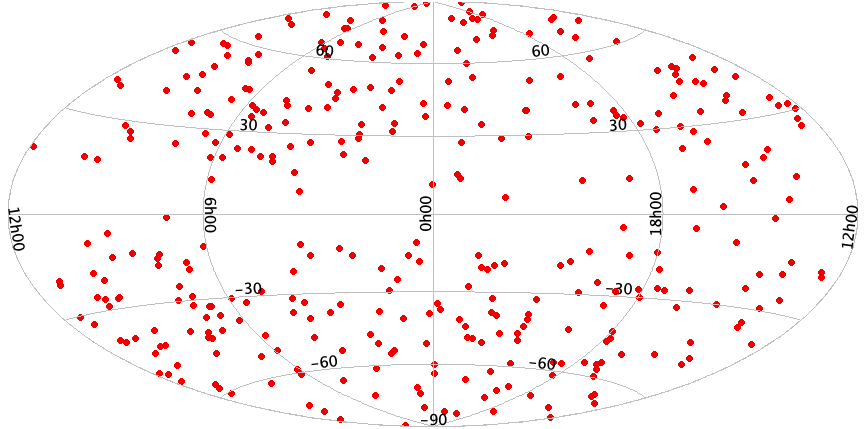}
    \caption{The distribution of all GRBs in the sample of \citet{horvath2015} in galactic coordinates. A clear extinction dependent bias can be seen.}
    \label{fig:AllSkyGRBs}
\end{figure}

I use two configurations of Monte Carlo sampling: one that assumes absolute isotropy and one that applies a correction for extinction and exposure effects. Extinction is known to limit the measurements of redshifts \citep{horvath2015}. In addition to extinction effects, the observations of gamma-ray bursts are limited by the sampling distribution of the telescopes used to discover them. Indeed, \citet{ukwatta2016} find that, using a kernel density estimation, the observed distribution of all GRBs that have measured redshifts closely follows a combined density map that considers extinction effects and the swift exposure map (using a slightly different data set). The extinction-based effects for the overall sample can clearly be seen in Fig. \ref{fig:AllSkyGRBs}.

Besides sampling effects, pure statistical considerations should be considered. The typical p-value significance threshold (0.05) applies to a singular test. Any trial that returns a significant value under a given p-value threshold is counted as an anisotropy (although the confidence of such an anisotropy is given by the amount the p-value varies from the pre-determined threshold). When applying statistical tests to many samples at once, a stricter p-value threshold should be used that limits the overall chance of a type I error (that is, the null-hypothesis is true but still rejected), particularly for such an important discovery as the HCB. The probability of at least one trial resulting in a false rejection of the null hypothesis is equivalent to:
\begin{equation}
    \alpha_{overall}=1-(1-\alpha_{ind})^{m}
    \label{eq:familywise}
\end{equation}
where $m$ is the number of trials, $\alpha_{ind}$ is the individual significance level, and $\alpha_{overall}$ is the overall significance level once all trials are taken into consideration. This is commonly called the family-wise error rate in the literature \citep{Feise2002}, and the specific procedure applied in this paper is commonly referred to as the Dunn-Sidak correction \citep{robertsokal1994}. If $\alpha_{overall}$ is desired to be 0.05, then $\alpha_{ind}$ must be 0.00512 for the nine group case. In other words, I have lowered the p-value to reduce the amount of errors made in the multi-sample study.

Yet, even this formulation doesn't consider the unknown number of trials that the authors of the original study conducted to find redshift ranges that returned a significant clustering. Admittedly, \citet{horvath2015} does use a large number of redshift ranges, with the suspect range in each redshift formulation returning a significant anisotropy (in certain tests, that is), but there are still potentially other redshift ranges not mentioned in the published paper that were used. Moreover, if one "extraneous" anisotropy exists, it is likely to show up in multiple redshift ranges. This effect of all tests essentially showing the same anisotropy explains why, under isotropic conditions, all tests could return a significant value for one redshift range if the family-wise error rate is set to the traditional significance value.

The distribution of GRBs in the sample is known to be effected by extinction. This can readily be seen in Fig. \ref{fig:AllSkyGRBs}. Moreover, the instruments observing GRBs have exposure effects. \citet{horvath2015} use the sky exposure map of Swift to demonstrate this, and both of these biases can also be approximated using empirical methods, as is done in this paper to quantitatively estimate effects of extinction and exposure on the derived significance levels and in \citet{horvath2015} to justify that extinction is not the cause of the HCB using a $\chi^2$ test.

To summarize, the main explanations for the differences in the p-value observed from a pure Monte Carlo simulation versus the unknown true probability are:
\begin{itemize}
    \item Extinction near the plane of the galaxy, clearly visible in Fig. \ref{fig:AllSkyGRBs}. While not affecting the overall distribution of GRBs, high extinction can limit measures of redshift
    \item Sky-exposure biases of gamma-ray telescopes
    \item The number of trials conducted in the nine-division configuration
    \item The unknown number of "test" trials that resulted in the original researchers' choice of redshift ranges
\end{itemize}

The first three points can be corrected for by considering extinction effects and applying the Dunn-Sidak correction. The final point can't be corrected for, thus p-values reported in my study likely still exaggerate the significance of the HCB, even when correcting for extinction and the nine division configuration.

\begin{figure}
	\includegraphics[width=\columnwidth]{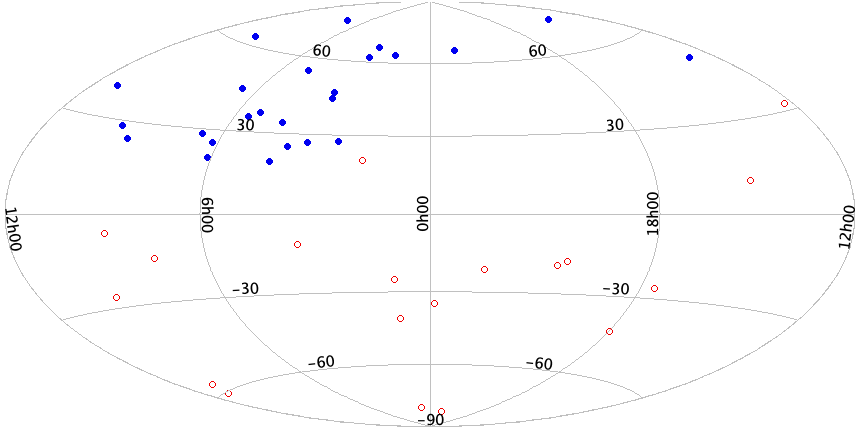}
    \caption{The distribution of GRBs in the 1.6 $<$ z $\leq$ 2.1 redshift range in galactic coordinates. The GRBs found to be within the maximum cluster for area \(0.1875 ×4\pi\) are colored blue and filled, while the rest of the sample is colored red and not filled.}
    \label{fig:GRBredshift}.
\end{figure}

\subsection{Point-Radius Method and Binomial Distribution}
\label{sec:prmethod} 
\citet{horvath2014} used what they call a ``bootstrap point-radius method'' to investigate anisotropies in the suspected redshift range of the HCB. I will hereby refer to this method as simply the point-radius method.
The procedure involves selecting the 44 bursts (in the updated 2015 data set) found within the suspect redshift range. Then, a large number of random points are generated on the celestial sphere. For each random point on the sphere, the distance to each of the 44 bursts is calculated, and the number of bursts within a certain angular radius is tabulated. Out of all the 10,000 random points used, the sample with the highest number of bursts within the predefined angular radius is reported. This procedure is repeated with multiple angular radii values, as well as with 44 random samples for comparison. The result of this procedure applied on the \citet{horvath2015} sample of GRBs is shown in Fig. \ref{fig:GRBredshift}. The point radius test does capture the visually apparent clustering in northern galactic latitudes.

\citet{horvath2014} estimated the probability that the observed number of bursts within an angular radius is not due to random chance using the principle that the number of bursts within a random angular radius should follow a binomial distribution. 

If a circle is defined with radius $\theta$ on the celestial sphere and samples are distributed isotropicaly, then the number of bursts within the circle should follow a binomial distribution:
\begin{equation}
    \binom{n}{k} p^{k}(1-p)^{n-k}
    \label{eq:binomial}
\end{equation}
where $p$, the probability of success, is given by the area of a spherical cap divided by the total surface area of the sphere (i.e. the solid angle), or:
\begin{equation}
    \frac{1}{2}(1-\cos{\theta})
    \label{eq:sphericalcap}
\end{equation}

\begin{figure}
    \centering
    \includegraphics[width=\columnwidth]{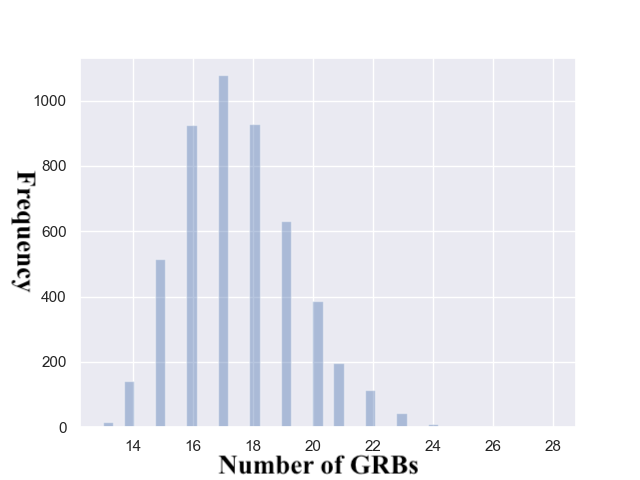}
    \caption{A histogram of GRBs within the specified angular radius for 5000 random isotropic samples for a point-radius test.}
    \label{figure:point-radius-random}
\end{figure}

The derivation of Eq. \ref{eq:binomial} assumes that the probabilities of success and failure for successive trials are independent of each other. The point-radius test, by finding the angular circle that contains the largest number of points, breaks that assumption of independence, and as such, a binomial distribution isn't appropriate. By finding the area with the most GRBs present, the number of bursts within the angular radius no longer follows a binomial distribution. This break from independence can be seen in Fig.~\ref{figure:point-radius-random}, which is generated using the procedure described below. 

If the angular radius is \(0.1875 ×4\pi\), then the probability of finding a single point in that circle is 0.1875. Using this as $p$ for Eq. \ref{eq:binomial}, the binomial distribution should have a mode of 8. However, the samples in Fig.~\ref{figure:point-radius-random} have a mode of 17.

As the correct distribution for the number of points in a circle containing the maximum number of points in an all-sky distribution cannot be found in the literature, it is helpful to find an empirical distribution using Monte Carlo sampling. \citet{horvath2014} found that the number of points within an observed radius is 25, and this equates to a probability equal to 0.0018 of being observed under isotropic conditions, using similar randomization techniques. This is hard to dispute without taking into account other effects. But when extinction is taken into account, this p-value will increase. 

To factor in extinction, the all-sky samples of GRBs is assumed to model the actual distribution of GRBs with extinction. Then, in the overall sample of GRBs (i.e. all of the simulation runs combined), the distribution of the all-sky sample of GRBs is matched. To properly model the all-sky sample of GRBs, I take the recommendations of \citet{horvath2014} of splitting up the sky into four different "zones" and for each zone assigning the proportion of GRBs found in the overall sample. These proportions are what are matched in the overall sample of GRBs in the extinciton simulations.

Taking the \(0.1875 ×4\pi\) area circle where \citet{horvath2015} found the largest deviation within the updated data set, I find that $P(X\geq x)$  is 0.002, where X is the probability distribution of random points with extinction taken into account within the angular radius and x is 25, the number of bursts found within the observed data set. Thus, extinction doesn't significantly change the resultant p-value.

My generated p-value is below traditional thresholds, but when taking the criteria of Sec. \ref{sec:statisticalconsiderations} into account, the p-value (0.002) almost reaches the significance threshold of 0.00512 to preserve the overall false positive rate of 0.05. Considering the unknown number of prior trials performed by the authors of the original study, this probability is likely even higher, although the unknown trials can't be quantified. Thus, the p-value that I think is the best one to quote in future studies, 0.002, still places moderate to strong doubt in the significance of the anisotropy. Moreover, this value rises when taking into account the updated data-set of Sec. \ref{sec:updated}.

\begin{figure}
    \centering
    \includegraphics[width=\columnwidth]{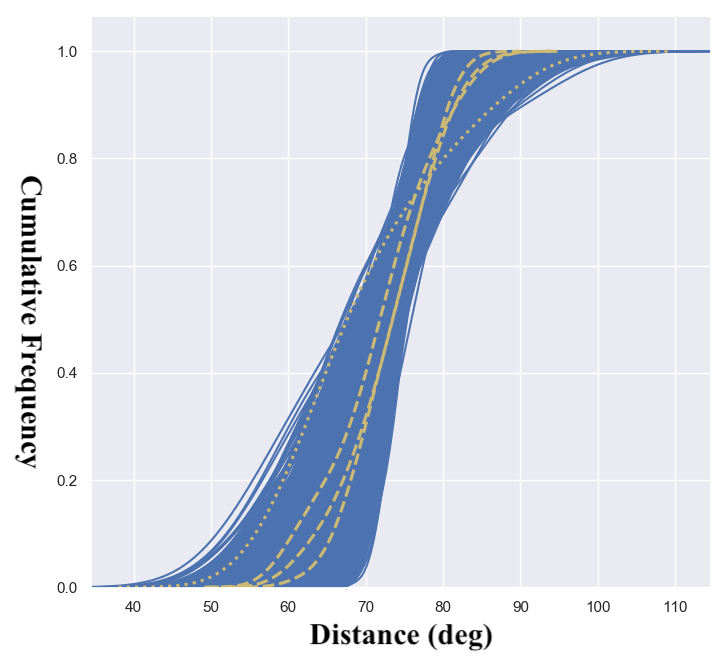}
    \caption{A cumulative distribution of the 31st k-th nearest neighbor distribution for the suspect range (dotted yellow), compared to 5000 Monte Carlo samples (blue). The three other groups in the four-group case are shown in dashed yellow for reference.}
    \label{fig:KDE_nneighbor}
\end{figure}
\subsection{K-th Nearest Neighbors Test}
\label{sec:kthnearestneighbor}
\citet{horvath2014} and \citet{horvath2015} also used a kth nearest neighbor test to demonstrate that the largest anisotropy was in \((1.61 \leq z < 2.68)\), and that the anisotropy was significant. By calculating the distance between each point and its kth nearest neighbor, they found that, when comparing the differences in the cumulative distributions of the kth nearest neighbor using a Kolmogorov-Smirnoff (KS) test of the suspect range and other ranges of redshift, the suspect redshift range always deviated significantly from the other ranges. Specifically, they found that the 31st nearest neighbor had the most significant deviation from what would be expected from isotropy.

It is best to consider the empirical meaning of the KS test. That is, the returned p-value might not represent the actual probability of drawing the sample from an isotropic distribution, while a Monte-Carlo derived distribution of significance values should give a more accurate estimation of the significance value. To implement this, I run a KS test of the GRBs in the suspect redshift range and randomly draw 44 GRBs assuming isotropy. This test is run 5000 times, and the mean K-statistic and p-value is calculated. The K-statistic and p-value are then compared to 5000 draws of a KS test of two isotropic samples. The results of these Monte Carlo simulations are shown in Fig. \ref{fig:KDE_nneighbor}. Although amongst the four ranges the suspect range is a clear outlier, there are still multiple Monte Carlo runs that return a higher vertical distance in cumulative distributions. I then repeat this procedure with the procedure for considering extinction effects of Sec. \ref{sec:prmethod} applied. I find that the probability of obtaining the observed K-statistic from the empirical Monte Carlo derived distribution of possible statistics for an isotropic sample is 0.0362, as 181 out of the 5,000 samples had a higher K-statistic (the results for p-values are naturally the same). This is in stark contrast to, for example, the derived p-value of $10^{-7}$, given as 0.9999999 in \citet{horvath2015} as the probability of the two distributions being different, where in their case groups one and two are being compared in the four group case.

\subsection{2D Kolmogorov-Smirnoff (KS) Test}
\label{2dks}
\citet{horvath2014} also uses a two dimensional KS test on the angular position data, based off of the formulation presented in \citet{peacock1983}. I implement this 2D KS test in Python using the "ndtest" package (\hyperlink{https://github.com/syrte/ndtest}{github.com/syrte/ndtest}). This package uses an updated version of the \citet{peacock1983} methodology, as found in \citet{fasano1987}. \citet{horvath2014} found the probability of the observed K-statistics by calculating an empirical distribution of expected K-statistics given two isotropic samples, and comparing this to the values of K-statistics when two observed samples were compared. 
Although this test was applied to the original dataset of \citet{horvath2014} -- and not in \citet{horvath2015} -- the test should give significant results as well for the dataset used in this study and in \citet{horvath2015} assuming that there is a significant anisotropy. 

\begin{figure}
    \centering
    \includegraphics[width=\columnwidth]{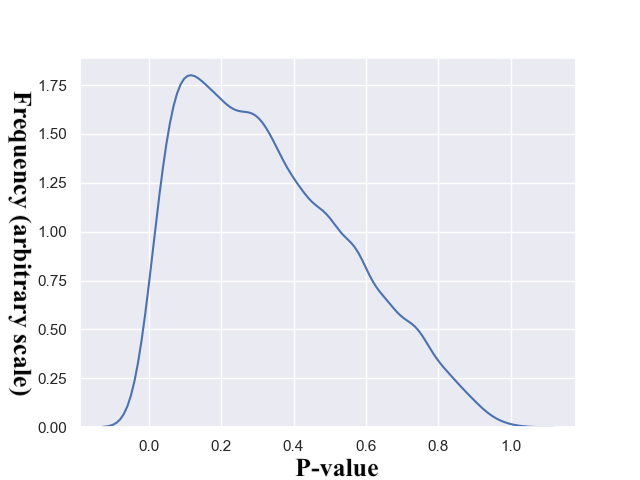}
    \caption{The distribution of p-values in a 2D-KS test of 5,000 Monte Carlo simulations as shown through a kernel density estimation.}
    \label{fig:2D-KS}
\end{figure}

To calculate the mean K-statistic and p-value, I compare the suspect redshift range distribution with that of 5,000 randomly generated isotropic samples. This generates an average "raw" p-value of 0.046, on par with the significance levels quoted in \citet{horvath2014}. This is equivalent to a K-statistic of 0.3456. When comparing this average K-statistic of the suspect redshift range to that of a K-statistic 5,000 random isotropic samples, higher significance values can be found in 0.0641 of the sample. In Fig. \ref{fig:2D-KS}, the distribution of p-values derived from Monte Carlo tests performed on the 2D-KS test show a nonuniform distribution, contrary to the uniform distribution that would be expected from an unbiased test. Yet, the "empirical" p-value (the value 0.0641 in this case) is surprisingly similar to the p-value without this correction (the "raw" p-value of 0.046), given the dramatic clustering of p-values in Fig.  \ref{fig:2D-KS} around small values. Nonetheless, it is still an increase in p-value, and well over the family-wise threshold according to Eq. \ref{eq:familywise}.
I use the same procedure as used throughout this paper to account for extinction biases. This lowers the p-value by 0.006, which I interpret to mean that extinction has no significant effect on this value. Running the simulation again with extinction, in fact, returns a slightly higher value, suggesting this is likely due to random variability in the derived statistics.

The reader might wonder why non-significant values are raised in this study, but not in \citet{horvath2014}. I postulate that there are two possible reasons for why this occurs:
\begin{itemize}
    \item The difference in formulation used (i.e. \citet{fasano1987} vs. \citet{peacock1983}). Although the two models have been shown to be similar in most cases, the \citet{fasano1987} model has greater accuracy with data that are not independent.
    \item The difference in sample used. \citet{horvath2014} used a smaller sample than the one used in this paper.
\end{itemize}

It is worth noting that the 2D KS test wasn't applied in \citet{horvath2015} for unspecified reasons. Nonetheless, this test still shows non-significant probability and, due to it being used in the original paper, is important to note.

\section{Updated Data Set}
\label{sec:updated}
The original study of \citet{horvath2014} used a sample of 283 GRBs with measured redshifts, with 31 of the GRBs lying within the suspect range. \citet{horvath2015} used an updated sample of 361 GRBs, of which 44 lie within the suspect range. In 2019, using all the GRBs located at \url{http://www.mpe.mpg.de/~jcg/grbgen.html} with redshifts, I collect an updated sample of 520 GRBs that have redshift measurements, with 75 lying within the suspect range.
The original tests of \citet{horvath2014} are performed on this data set as well, with the results summarized here (with the results of the tests performed on the data set of \citet{horvath2015} for comparison):
\begin{center}
 \begin{tabular}{||c c c c||} 
 \hline
 Sample & Point-Radius test & K test & 2D KS test \\ [0.5ex] 
 \hline\hline
 2015 sample & 0.002 & 0.0362 & 0.064 \\ 
 \hline
 2019 sample & 0.0294 & 0.054 & 0.2232 \\
 \hline
\end{tabular}
\end{center}

Each number in the table represents the probability of finding the observed statistic for each test assuming an isotropic distribution and considering extinction effects using the correction described in \ref{sec:prmethod}. From Sec. \ref{sec:statisticalconsiderations}, the threshold of significance for the HCB being an anisotropy is taken to be 0.00512. It follows that any individual p-values from tests in the table that are above 0.00512 are considered to suggest non-significance. The marked increase in p-value in the updated sample suggests that the low probability found in Sec. \ref{sec:significance} is likely a coincidence of the sample.

\section{Discussion and Conclusions}
The HCB, if it exists, would be the largest known structure in the universe. But the existence of the structure has come under scrutiny \citep{ukwatta2016}. In this paper, I have gone back to the original methods used to discover the HCB in \citet{horvath2014} to see if the results they reported exaggerated statistical significance, and I apply the original statistical tests to a larger data set.

I go through each test that \citet{horvath2014} used. For the point-radius test of Sec. \ref{sec:prmethod}, a binomial distribution is an invalid way to calculate a p-value. Finding the maximum number of GRBs in the observed radius introduces a bias. Monte Carlo simulations can be drawn to derive a p-value instead, eliminating the need to calculate a theoretical distribution for the point-radius test. In the case of the Kth nearest neighbor test of Sec. \ref{sec:kthnearestneighbor} and 2D KS test of Sec. \ref{2dks}, Monte Carlo simulations drawn from an isotropic distribution demonstrated a non-uniform distribution of p-values, which could consequently bias the p-value derived in the case of the true sample towards abnormally low values. These biases were corrected for by deriving p-values from a comparison of the observed sample to Monte Carlo simulations.

If the family-wise error rate of finding the HCB is to be maintained at 0.05, using the Dunn-Sidak correction suggests that any one p-value should be below 0.00512 to consider the result significant. This Dunn-Sidak correction only accounts for the known amount of trials in \citet{horvath2014}: the nine redshift ranges. It is unknown how many tests and varying redshit ranges were used before finding the original significant results of \citet{horvath2014}. If the true number of trials was accounted for, the probability would likely rise above the numbers reported in my analysis.

I then apply these tests to an updated sample that I compile as of August 2019. This updated sample has an almost two-fold increase in the number of GRBs present in the suspect redshift range.

I find that the p-values returned by each statistical test are non-significant when considering the Dunn-Sidak correction except for the point-radius test from the sample of \citet{horvath2015}. I argue that this doesn't suggest significance by itself, since in the more comprehensive sample that I compile, the point radius test returns non-significance.

As would be expected if the distribution of GRBs was isotropic, a larger sample size increases the p-values derived in various tests. As time progresses, more GRBs will have measured redshift and time will tell if ultimately the p-values increase into the realm of definitive isotropy or decrease to the point that the HCB definitively exists.

\citet{ukwatta2016} have also found no significant anisotropy in the suspect range using different methods and a slightly different data set. In fact, their kernel density estimation has shown that the suspect region's anisotropy matches almost perfectly with a combined instrument selection and extinction density map.

Although the existence of the HCB has always been suspect, now even the methods applied in the papers purporting its existence have been shown to not actually suggest any significant anisotropies. With this in consideration, the existence of the HCB is further in doubt.

Assuming the HCB doesn't exist, the current $\Lambda$CDM cosmological paradigm faces one less conundrum in its explanation of the large-scale structure of the universe. Some have argued that Hercules-Corona Borealis Great Wall doesn't even challenge the $\Lambda$CDM paradigm. The HCB is larger than the commonly cited \citet{yadav2010} value of 260 Mpc $h^{-1}$ for the scale of homogeneity, but \citet{li2015} has argued that a larger scale of homogeneity should be used for gamma-ray bursts, around 8600 Mpc $h^{-1}$, which would mean that the HCB doesn't violate the $\Lambda$-CDM model. Other explanations for the existence of the HCB besides an structural anisotropy have also been presented: \citet{balazas2018} argues that an anisotropy found in GRBs is more likely to represent an increase in star formation rate than a physical structure. Although there are certainly other structures larger than the theoretical maximum anisotropy size of \citet{yadav2010} (but not the \citet{li2015} value of 8600), the HCB is by far the largest claimed, with approximately twice the size of the second largest structure. Given the mounting doubt cast over the existence of the HCB over time, future works should scrutinize the purported anisotropy in other structures that possibly violate the $\Lambda$-CDM model. For example, multiple structures have been found in the structure of quasars\footnote{\citet{Nadathur2013} criticizes this finding, pointing out that such anisotropic signals occur randomly in simulations.} of an approximate size of 350 Mpc of \citet{clowes2011} and the even larger "Huge large quasar group" (commonly abbreviated Huge-LQG) of approximate size 500 Mpc \citep{Clowes2013}.

\section*{Acknowledgements}





\bibliographystyle{mnras}
\bibliography{refs}




\bsp	
\label{lastpage}
\end{document}